\documentclass[aps,pra,twocolumn,noshowpacs,showkeys]{revtex4-1}
\usepackage{xcolor, graphicx, ulem}
\usepackage{amsmath, amssymb}
\usepackage[colorlinks=true,urlcolor=blue,citecolor=blue,linkcolor=blue]{hyperref}

\begin{document}

\title{Time-dependent nonlinear Jaynes-Cummings dynamics of a trapped ion}

\author{F. Krumm}\email{fabian.krumm@uni-rostock.de}\affiliation{Arbeitsgruppe Theoretische Quantenoptik, Institut f\"ur Physik, Universit\"at Rostock, D-18059 Rostock, Germany}
\author{W. Vogel}\affiliation{Arbeitsgruppe Theoretische Quantenoptik, Institut f\"ur Physik, Universit\"at Rostock, D-18059 Rostock, Germany}

\begin{abstract} 
In quantum interaction problems with explicitly time-dependent interaction Hamiltonians, the time ordering plays a crucial role for describing the quantum evolution of the system under consideration. 
In such complex scenarios, exact solutions of the dynamics are rarely available. Here we study the nonlinear vibronic dynamics of a trapped ion, driven in the resolved sideband regime with some small frequency mismatch. 
By describing the pump field in a quantized manner, we are able to derive exact solutions for the dynamics of the system. This eventually allows us to provide analytical solutions for various types of time-dependent quantities. 
In particular, we study in some detail the electronic and the motional quantum dynamics of the ion, as well as the time evolution of the nonclassicality of the motional quantum state. 
\end{abstract}


\date{\today}
\maketitle

\section{Introduction}\label{Sec:Introduction}

	The verification and quantification of nonclassical effects, that is, phenomena which cannot be explained by Maxwell's equations, is a main concern of theoretical and experimental quantum optics.
	Many of those effects, like squeezing~\cite{W83,Slusher,Wu,Va08,Va16}, entanglement~\cite{EPR35,S35}, and photon antibunching~\cite{KDM77}, were intensively investigated over many decades.
	However, there are effects beyond this set, like, for example, anomalous quantum correlations~\cite{V91,KVMKH17,GR09}, which arise from the violation of field-intensity inequalities.
	In such and related scenarios a subject of interest is the investigation of the interplay of free fields and fields which are attributed to sources, which play an important role in the theory of spectral filtering of light \cite{KVW86,Cresser87}.
	The relationships between field correlation functions of free-field and source-field operators were, for example, considered in Refs.~\cite{KVW87,Stokes17}.
	Hence, the treatment of a physical system containing contributions from both kinds of fields is an interesting aspect to be studied, especially when the corresponding dynamics is exactly solvable.
	A suitable model for this purpose is the Jaynes-Cummings model, which contains not only free-field parts but a source-attributed part as well. In the following we will briefly reconsider its history and possible areas of application.
	
	When the Jaynes-Cummings model was proposed in 1963~\cite{JC63,P63}, its practical relevance was doubted, as it describes an idealized scenario of the resonant interaction of a two-level system with only a single radiation mode.	
	However, in the 1980s the model's importance was vastly enhanced, since, due to technical progress, it was possible to experimentally prove many of its predictions~\cite{H84,HR85,MWM85,RWK87}.
	Remarkably, despite its simplicity, the Jaynes-Cummings model exhibits plenty of physical effects, e.g. Rabi oscillations~\cite{R36,R37,ERD03}, collapse and revivals~\cite{RWK87,E80,N81}, squeezing~\cite{K88,H89}, 
	atom-field entanglement~\cite{S91,F99,B01}, antibunching~\cite{C86,H05,D07}, and nonclassical states such as Schr\"odinger cat~\cite{B92,GZ96} and Fock states~\cite{S89,W99,BV01}.
	Initially intended for describing the interaction of a single atom with a single radiation mode, the Jaynes-Cummings model could be applied to a variety of physical scenarios.
	Examples are Cooper-pair boxes~\cite{I03,W04}, "flux" qubits~\cite{C04}, and Josephson junctions~\cite{H96,G98,S04}.
	It can also be applied in solid-state systems to describe the (strong) coupling of qubits to a cavity mode, for example in quantum dots~\cite{M04,K10,B13,F17} or superconducting circuits~\cite{F08,F12,N10}.
	Another recent application of the Jaynes-Cummings model is the description of Rydberg-blockaded atomic ensembles~\cite{K16,L17}.
 	
        The Jaynes-Cummings model also became relevant for the vibronic dynamics of trapped ions, where the quantized mode of the electromagnetic field is replaced by the quantized center-of-mass-motion of the ion~\cite{B92a,B92b,CBZ94}. 
	Later on, a nonlinear Jaynes-Cummings model (NJCM) was introduced~\cite{V95}, which describes the dynamics of a trapped ion beyond the standard Lamb-Dicke regime~\cite{M96,L03}.
	The motional degrees of freedom are coupled to the electronic states of the ion by a classical pump field in the resolved sideband regime.
	On this basis it became possible to generate many motional states for trapped ions, such as Fock states, squeezed states~\cite{C93,M96}, even and odd coherent states~\cite{F96}, nonlinear coherent states~\cite{MV96}, pair coherent states~\cite{GS96a}, superpositions of the latter~\cite{GS96b}, SU(1,1) intelligent states~\cite{G97}, Schr\"odinger cat states~\cite{MM96,G96}, 
	entangled coherent states~\cite{G96}, and generalized Kerr-type states~\cite{WV97}.
	As for the standard Jaynes-Cummings Hamiltonian (see, for example,~\cite{L08}), the trapped-ion dynamics based on the nonlinear Jaynes-Cummings model was also considered beyond the rotating-wave approximation~\cite{MC2012,P15,M16,C17}.

	In the present paper we study the vibronic nonlinear Jaynes-Cummings model, when the classical driving laser field is slightly detuned from the $k$-th sideband.
	Such a mismatch yields an explicitly time-dependent Hamiltonian in the Schr\"odinger picture, the corresponding dynamics of which is not easily solved due to the relevance of time-ordering effects.
	We will demonstrate that these difficulties can be resolved by extending the Hilbert space of the problem to include the driving field in the quantum description. On this basis, 
	the full dynamics will become exactly solvable. This renders it possible to study sophisticated problems of explicitly time-dependent dynamics on the basis of the exactly solvable extended problem. 
	This yields deeper insight into the yet rarely studied quantum dynamics in cases when explicitly time-dependent Hamiltonians and the resulting time-ordering prescriptions are relevant. 
	Also the quantum effects and the nonclassical correlation properties of the system can be studied by this method in great depth.

	The paper is structured as follows. 
	Section \ref{Sec:JC} introduces the Hamiltonian used in this paper and briefly discusses its physical meaning as well as time-ordering effects. 
	In Sec. \ref{Sec:SOL}, we solve the dynamics using the eigenstates of the generalized Hamiltonian.
        Afterwards, in Sec. \ref{Sec:Pfunc}, we use the regularized Glauber-Sudarshan $P$~function to study the nonclassical evolution of the motional quantum properties of the ion.
	Finally, a summary and some conclusions follow in Sec. \ref{Sec:Conclusions}.
	
	\section{Explicitly time-dependent nonlinear Jaynes-Cummings model}\label{Sec:JC}

	The NJCM for the vibronic coupling between the electronic and motional degrees of freedom of a trapped ion was introduced for the situation of the exactly resonant interaction of a laser field 
	with the $k$-th vibronic sideband of the ion~\cite{V95,VW06}. 
	This interaction Hamiltonian can be exactly diagonalized. 
	It was experimentally demonstrated that it properly describes the dynamics of a trapped ion for the case of $k=1$~\cite{M96}. 
	In the present paper we are interested in more sophisticated time-dependent quantum phenomena of such a system. 
	For this reason, in a first step we generalize the NJCM to allow for the explicitly time-dependent dynamics. 
	In this case, however, an exact solution of the problem seems to be not feasible and numerical solutions are required.
	
         \subsection{Explicitly time-dependent Hamiltonian}
	 Let us start with the following Hamiltonian, describing an ion, trapped in a harmonic trap potential, interacting with a classical laser field (see~\cite{V95} and Chap. 13 of~\cite{VW06}):
	  \begin{align}
	 \hat H(t)=& \underbrace{\left(  \hbar \nu \hat a^\dag \hat a + \hbar \omega_{21} \hat A_{22}\right)}_{=\hat H_0}   \label{Eq:StartH0}  \\
	 &+ \underbrace{\left( \hbar \kappa| \beta_\text{cl} | e^{-i \omega_L t} \hat A_{21}\hat g( \eta(\hat a + \hat a^\dag)) + \text{H.c.}  \right)}_{=\hat H_\text{int}(t)}.  \label{Eq:StartH}
	 \end{align}
	 $\hat H_0$ describes the free motion of the vibrational center-of-mass and electronic degrees of freedom of the two-level ion, with the 
	 vibrational frequency $\nu$ and the electronic transition frequency $\omega_{21}=\omega_2-\omega_1$.
	 The laser is assumed to be monochromatic and quasiresonant with the electronic $|1 \rangle \leftrightarrow |2 \rangle$ transition, $\omega_L \approx \omega_{21}$, and 
	 to have only one non-vanishing wave-vector component, such that only one motional degree of freedom appears in the interaction term.
	 The complex amplitude  $\beta_\text{cl}$ describes the pump laser.
	 The operators $\hat a^\dag$  ($\hat a$) are the creation (annihilation) operators of the vibrational frequency $\nu$.
	 The electronic flip operators $\hat A_{ij}=|i\rangle \langle j|$ ($i,j=1,2$) describe the atomic $|j\rangle \rightarrow |i\rangle$ transitions,
	 $\kappa$ is a projection of the electric-dipole matrix element on the direction of the electrical field, and $g( \eta(\hat a + \hat a^\dag))$ describes the mode structure of the pump laser at the operator-valued position of the ion.
	 For a standing wave it reads as
	 \begin{align}
	  &g( \eta(\hat a + \hat a^\dag)) = \cos [  \eta(\hat a + \hat a^\dag ) + \Delta \phi   ],
	 \end{align}
	 where $\Delta \phi$ defines the relative position of the trap potential to the laser wave.
	 The Lamp-Dicke parameter $\eta$ describes the effects of momentum transfer on the atomic wave packet due to recoil effects.

	 Applying the Baker-Campbell-Hausdorff formula in $\hat g$ together with a power series expansion, we get
	 \begin{align}
	 \hat g( \eta(\hat a + \hat a^\dag)) = \frac{1}{2} e^{i \Delta \phi -\eta^2/2} \sum_{l,m=0}^\infty \frac{(i \eta)^{l+m}}{l!m!} \hat a^{\dag l} \hat a^m +\text{H.c.}   \label{Eq:Expansiong}
	 \end{align}
	 The interaction Hamiltonian in the interaction picture (indicated by the tilde) reads as
	 \begin{align}
	 &\hat{\tilde H}_\text{int}(t) = \frac{1}{2} \hbar \kappa| \beta_\text{cl}|\hat A_{21} e^{ -\eta^2/2}  \nonumber  \\ 
	 &\times \sum_{l,m=0}^\infty   \frac{\hat a^{\dag l} \hat a^m }{l!m!} e^{-i[\omega_L-\omega_{21}+(m-l)\nu]t} \nonumber  \\
	 &\times \Big\{ e^{i \Delta \phi} (i \eta)^{l+m}  +  e^{-i \Delta \phi}  (-i \eta)^{l+m}\Big\} + \text{H.c.}    \label{Eq:Hinteraction1-intpic}
	 \end{align}
	 If the laser is exactly resonant to the $k$-th vibrational sideband, this yields the exactly solvable nonlinear Jaynes-Cummings interaction~\cite{V95,VW06}.

\begin{figure}[t]
	\centering
	\includegraphics*[width=0.4\textwidth]{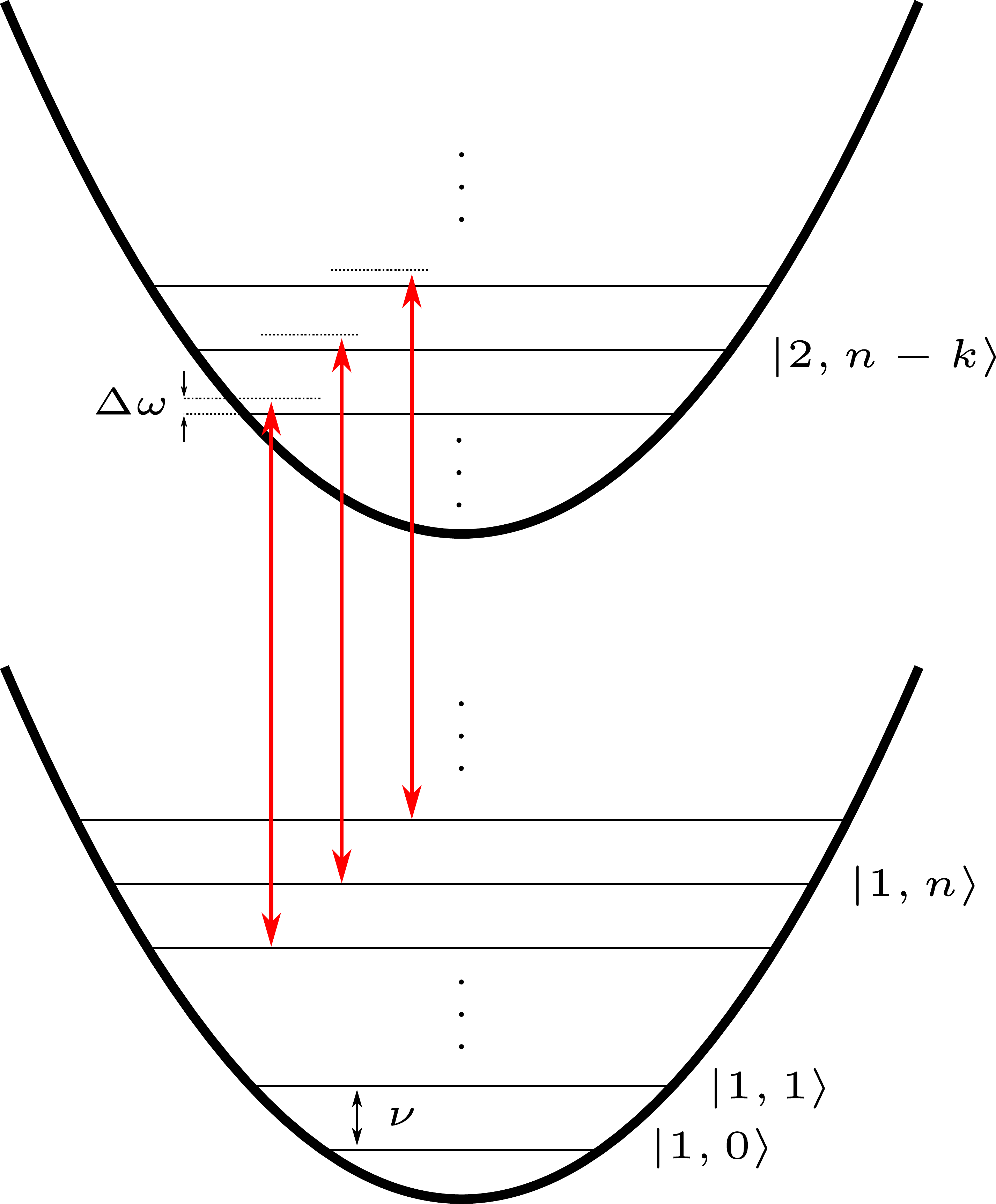}
	\caption{ Scheme of the dynamics described by the interaction Hamiltonian~\eqref{Eq:Hinteraction2-intpic}. 
	The two electronic states, ground state $|1\rangle$ (lower potential-energy surface) and excited state $|2 \rangle$ (upper potential-energy surface), are separated by the electric transition frequency $\omega_{21}=\omega_2-\omega_1$. 
	Since the (harmonic) trap potential is not influenced by the ion dynamics, the energy surfaces are neither displaced, nor distorted and the vibronic levels are 
	separated equidistantly by $\nu$. The laser frequency $\omega_L=\omega_{21} - k \nu + \Delta \omega$ (red arrows) is not in exact resonance with the $|1,n\rangle \leftrightarrow |2,n-k \rangle$ transition but slightly detuned by $\Delta \omega$.
	}\label{Fig:JCmodel}
	\end{figure}

	 For the purpose of the present paper, we are interested in the situation when the laser is slightly detuned from the $k$-th sideband:
	 \begin{equation}
	  \omega_L=\omega_{21}-k \nu + \Delta \omega, \label{Eq:mismatch-cond}
	 \end{equation}
	 with $\Delta \omega \ll \nu$.
	 We still assume that the ion is in the resolved sideband limit, i.e.,  we can resolve the single sidebands very well.
	 This means that the linewidths of the vibronic transitions and the coupling strength $|\kappa|$ are small compared to the vibrational frequency $\nu$.
	 In this case one only excites vibronic transitions which are quasiresonant according to the condition in Eq.~\eqref{Eq:mismatch-cond}, which are
	 the $|1,n \rangle \leftrightarrow |2,n-k\rangle$ transitions for $k\geq 0$.
	 Hence, we perform a vibrational rotating-wave approximation, 
	 \begin{equation}
	   e^{\mp in \nu t} =0 \quad  \forall n \neq 0,
	 \end{equation}
	 in Eq.~\eqref{Eq:Hinteraction1-intpic}. 
	 This yields the interaction Hamiltonian 
	 \begin{align}
	 &\hat{\tilde H}_\text{int} (t)= \hbar \kappa| \beta_\text{cl}|e^{- i\Delta \omega t} \hat A_{21}  \hat f_k(\hat a^\dag \hat a;\eta) \hat a^k+ \text{H.c.},   \label{Eq:Hinteraction2-intpic}
	 \end{align}
	 where
	 \begin{align}
	& \hat f_k(\hat a^\dag \hat a;\eta)= \frac{1}{2} e^{i \Delta \phi -\eta^2/2} \sum_{l=0}^\infty \frac{(i \eta)^{2l+k}}{l! (l+k)!} \hat a^{\dag l} \hat a^l + \text{H.c.}   \nonumber \\
	&=\frac{1}{2} e^{i \Delta \phi -\eta^2/2}  \sum_{n=0}^\infty |n\rangle \langle n | \frac{(i \eta)^k n!}{(n+k)!} L_n^{(k)} (\eta^2) + \text{H.c.}  , \label{Eq:def-f}
	\end{align}
	with $L_n^{(k)} $ denoting the generalized Laguerre polynomials.
	The Hamiltonian~\eqref{Eq:Hinteraction2-intpic} describes the nonlinear $k$-th sideband coupling of the vibrational mode and the electronic transition, $|1,n \rangle \leftrightarrow |2,n-k\rangle$ (see Fig.~\ref{Fig:JCmodel}). 
	It is important that this nonlinear interaction is explicitly time dependent, as long as $\Delta \omega \neq 0$.

	\subsection{Solution of the time-dependent interaction and time-ordering effects}
	\label{SubSec:TimeOrdering}
	The most general solution of the time evolution of a quantum system, described by its Hamiltonian $\hat H(t)$, is given by its time-evolution operator:
	\begin{align}
	 \hat U(t) = \mathcal{T} \exp \left\{ \frac{-i}{\hbar} \int_{t_0}^t \hat H(t') dt'\right\}, \label{Eq:U1}
	\end{align}
	where ${\mathcal{T}}$ denotes the time-ordering prescription.
	The latter accounts for the temporal order of the Hamiltonians with different time arguments contained in the exponential function.
	If the Hamiltonian, however, is not explicitly time dependent or commuting with itself at different times, the ordering symbol ${\mathcal{T}}$ becomes superfluous and the standard exponential power series of $\hat H$ is recovered.

        Alternatively, the representation~\eqref{Eq:U1} can also be given in the form of the Magnus expansion~\cite{M54,B09}:
	\begin{align}
	 \hat U(t)= \exp \left\{  \frac{-i}{\hbar} \int_{t_0}^t \hat H(t') dt' + \sum_{n=1}^\infty \hat \Omega_n(t) \right\}, \label{Eq:U-Magnus}
	\end{align}
	which is  unitary in each order $n$, with $n=1, \dots, \infty$.
	Herein, the contributions of $\hat \Omega_n(t)$ are referred to as \textit{time-ordering effects} or \textit{time-ordering corrections}~\cite{CB13,QS14,QS15,QS16,KSV16}.
	However, the $\hat \Omega_{ n } (t)$ contain multiple integrals of nested commutators of the Hamiltonians at different  times which are, especially for higher orders, difficult to handle.
	A possibility to circumvent this problem was presented in Ref.~\cite{AF11}, where only one commutator needs to be evaluated.
	However, in this representation a needed diagonalization of the operator-valued problem is not trivial, as the different orders of the expansion do not necessarily possess a common eigenbasis.
	For certain physical models and regimes, the time-ordering symbol $\mathcal{T}$ in~\eqref{Eq:U1} can be neglected, for example for parametric down-conversion with not too high pump powers~\cite{CB13}.
	Hence, let us begin to study the influence of time-ordering effects on the dynamics described by the Hamiltonian in Eq.~\eqref{Eq:Hinteraction2-intpic}.
	
	For this purpose we use the open-source software package \textsc{qutip}~\cite{Qutip1,Qutip2} in \textsc{python} to obtain numerically the time-ordered solutions based on Eq.~\eqref{Eq:U1} together with the Hamiltonian~\eqref{Eq:Hinteraction2-intpic}.
        To visualize the effects of time ordering, we compare the solutions with those when the time ordering is discarded,
        $\hat{\tilde U} \to \hat{\tilde U}'$, 
	\begin{align}
	\hat{\tilde U}'(t)= \exp \left\{  \frac{-i}{\hbar} \int_{t_0}^t \hat{\tilde{H}}_\text{int}(t') dt'  \right\}. \label{Eq:U-no-timeordering}
	\end{align}
	In this case the integral can be directly evaluated, 
	\begin{align}
	\label{Eq:U-no-timeordering1}
	 &\int_{t_0}^t  \hat{\tilde{H}}_\text{int}(t')  dt' =  \hbar \kappa| \beta_\text{cl}|  \nonumber \\
	 &\times \frac{i}{\Delta \omega} \left( e^{-i\Delta \omega t}- e^{-i\Delta \omega t_0}\right) \hat A_{21}  \hat f_k(\hat a^\dag \hat a;\eta) \hat a^k + \text{H.c.}
	\end{align}
	For convenience we introduce the dimensionless quantities
	\begin{align}
	\label{Eq:dimquant1}
	 &r :=\frac{\Delta \omega}{|\kappa \beta_\text{cl}|},  \\
	 & \tau_{(0)} := |\kappa \beta_\text{cl} | t_{(0)},  \label{Eq:dimquant2}
	\end{align}
	such that
	\begin{align}
	 \Delta \omega t_{(0)} =  r \tau_{(0)}.
	\end{align}
	Furthermore, we define $h(\tau) := i \left(e^{-i r\tau}- e^{-i  r\tau_0} \right)$.
	Hence, we have
	\begin{align}
	 & \hat{\tilde U}'(t) \nonumber \\
	 & =\exp \left\{ -i \left[ \frac{h(\tau)}{r}  e^{i \arg (\kappa) } \hat A_{21}  \hat f_k(\hat a^\dag \hat a;\eta) \hat a^k + \text{H.c.} \right]  \right\}  \label{Eq:U-no-timeordering2}
	\end{align}
	and we assume $\arg (\kappa) =0$ from now on.

	The eigenstates of the integrated Hamiltonian~\eqref{Eq:U-no-timeordering1} read as	
	\begin{align}
	 | \psi_n^\pm \rangle = c_n^\pm ( |2,n\rangle + \alpha_n^\pm |1,n+k \rangle ),
	\end{align}
	with $|i,n\rangle$ denoting the electronic ($i=1,2$) and motional ($n=0,1,2,\dots$) excitations (see, e.g., Chap. 12 of Ref.~\cite{VW06}).
	Due to normalization we find immediately $c_n^\pm  = \frac{1}{\sqrt{1+|\alpha_n^\pm|^2}}$.
	These states $ | \psi_n^\pm \rangle$ are often referred to as ``dressed states.''
	Solving 
	\begin{align}
	 \left[ \frac{h(\tau)}{r}  \hat A_{21}  \hat f_k(\hat a^\dag \hat a;\eta) \hat a^k + \text{H.c.} \right]  | \psi_n^\pm \rangle \overset{!}{=} \omega_n^\pm  | \psi_n^\pm \rangle 
	\end{align}
	yields the parameters
	\begin{align}
	&\alpha_n^\pm =\pm e^{-i \arg (f_k(n;\eta)h(\tau) ) }, \quad c_n^\pm =\frac{1}{\sqrt{2}} \nonumber \\
	&\omega_n^\pm = \pm \frac{|f_k(n;\eta) h(\tau)|}{r}\sqrt{\frac{(n+k)!}{n!}} \equiv \omega_n^\pm (\tau) ,
	\end{align} 
	where $f_k(n;\eta)=\langle n | \hat f_k(\hat a^\dag \hat a ;\eta) | n \rangle$, [see Eq.~\eqref{Eq:def-f}].
	The completeness relation of these states reads
	\begin{align}
	 \hat 1= \sum_{\sigma=\pm} \sum_{n=0}^\infty |\psi_n^\sigma \rangle \langle \psi_n^\sigma | + \sum_{q=0}^{k-1} |1,q\rangle \langle 1,q|.
	\end{align}
	This yields the time evolution operator~\eqref{Eq:U-no-timeordering2} in the form 
	\begin{align}
	  \hat{\tilde U}'(\tau) = \sum_{\sigma=\pm} \sum_{n=0}^\infty e^{-i \omega_n^\sigma (\tau)}|\psi_n^\sigma \rangle \langle \psi_n^\sigma |,
	\end{align}
	since the $\sum_{q=0}^{k-1} |1,q\rangle \langle 1,q|$ part cancels, as $\hat a^k |q\rangle=0$ for $q < k$.
	
\begin{figure}[t]
	\centering
	\includegraphics*[width=8.0cm]{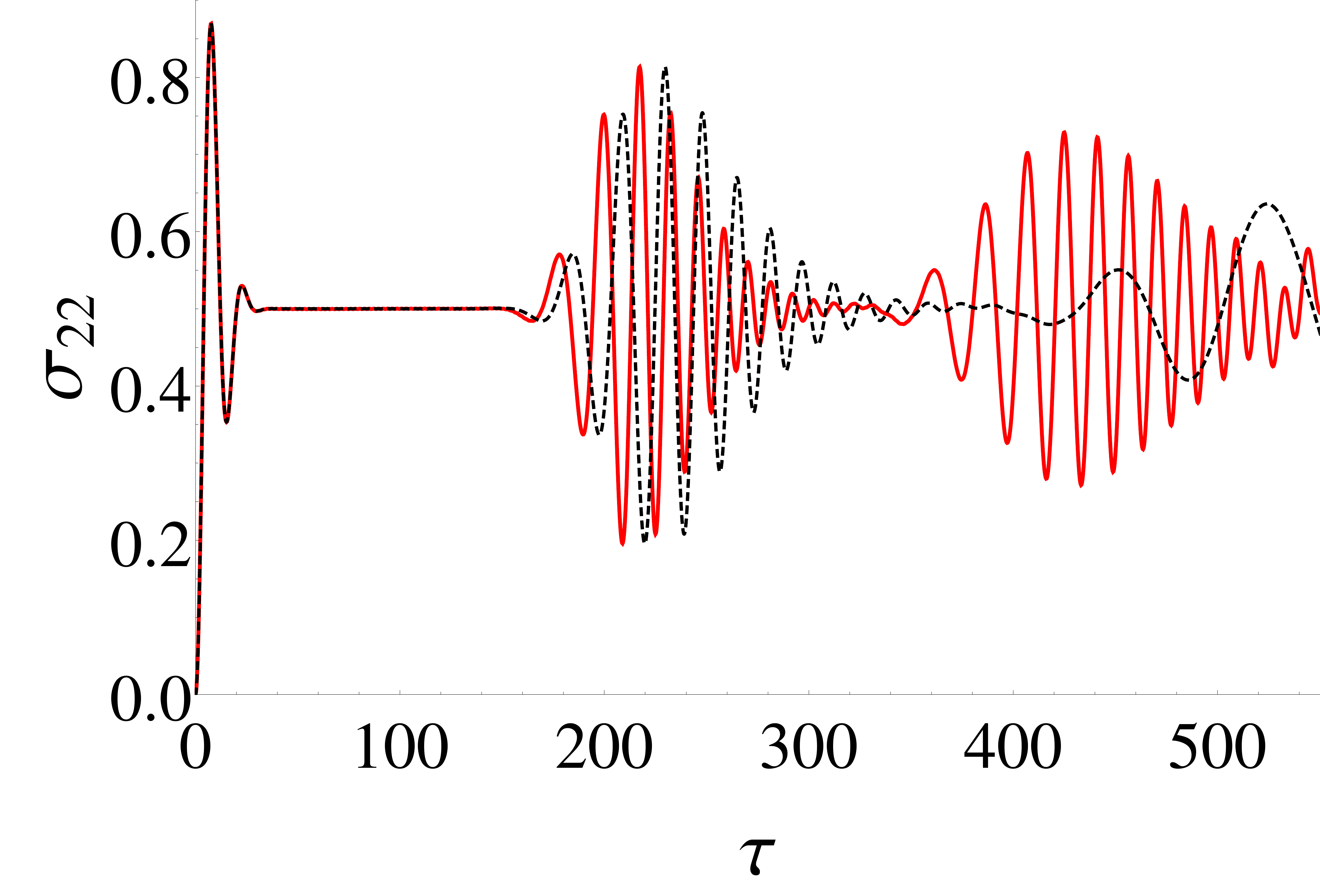}
	\caption{
	Influence of the time ordering on the population dynamics of the excited electronic state of the ion.
	The curves represent the numerical solution $\sigma_{22} (\tau)$ (solid, red line) for a small mismatch $r=0.005$ and, for the same $r$, the analytical solution without time ordering, $\sigma_{22} '(\tau)$ (dashed, black line).
	The motional degree of freedom is initially prepared in the coherent state $|\alpha_0\rangle$.
	Parameters: $\alpha_0=\sqrt{12}$, $k=2$, $\Delta \phi=0$, and $\eta=0.2$. 
	}\label{Fig:timeordering}
	\end{figure}

	For further investigations let us consider the population probability of the excited electronic state, which was studied only for $\Delta \omega=0$ in Ref.~\cite{V95}:
	\begin{align}
	 & \sigma' _{22}(\tau) \nonumber \\
	 &= \sum_{n=0}^\infty \langle 2,n|  \hat{\tilde U}'(\tau)  \hat U_0(\tau) \hat \rho (0) \hat U_0(\tau)^\dag  \hat{\tilde U}'(\tau)^\dag  |2,n\rangle,
	\end{align}
	which is given now in dependence on the scaled time $\tau$.
	For the visualization we chose the input state $|1,\alpha_0 \rangle$ at $ \tau_0=0$. 
	The atom is initially prepared in the electronic ground state and the motional state of the ion is a coherent state.
	Details concerning the coherent-state preparation of the motional state of the ion can be found in Refs.~\cite{M96,Wine90}.
	This eventually yields
	\begin{align}
	 & \sigma '_{22}(\tau) = \frac{1}{4} \sum_{n=0}^\infty \sum_{\sigma,\sigma'=\pm} e^{i[ \omega_n^\sigma (\tau)-\omega_n^{\sigma'}(\tau)  ]} \nonumber \\
	 &\times (\alpha_n^{\sigma'})^\ast \alpha_n^\sigma \frac{|\alpha_0|^{2n+2k}}{(n+k)!} e^{-|\alpha_0|^2}.
	\end{align}
	Note that the $\hat U_0(\tau)$ contributions cancel each other. 
	The temporal evolution of $\sigma '_{22}(\tau) $ is depicted in Fig.~\ref{Fig:timeordering}.
	The correct numerical solution significantly differs from the  analytical one without time ordering.
	That is, neglecting the time-ordering effects, even for a very small frequency mismatch $r$, strongly falsifies the electronic population dynamics.
	Hence,  the ${\mathcal T}$ ordering plays an important role and it must not be omitted.
	
	\section{Nonlinear Jaynes-Cummings Model with quantized pump}
	\label{Sec:SOL}
        In this section we will overcome the shortcoming of the nonlinear Jaynes-Cummings model with frequency mismatch by quantization of the pump field. 
        In practice, this can be realized by placing the trapped ion in a high-$Q$ cavity. 
        Now a mode of the quantized cavity field pumps the vibronic transition.  
        We will see that this extension of the Hilbert space allows us to exactly solve the full interaction problem. 
        This opens new possibilities to study problems underlying time ordering analytically, which are not solvable in a semiclassical approach. 
	
	\subsection{Quantization of the pump field}
	\label{Subsec:QuantPump}
	As shown above, the explicit time dependence of the Hamiltonian~\eqref{Eq:Hinteraction2-intpic} prevents an analytical solution of the dynamics as we cannot discard---even approximately---the time-ordering effects.
	Hence, our aim is to eliminate the time dependence in the Hamiltonian $\hat H(t)$. 
	For this purpose we return to the Hamiltonian~\eqref{Eq:StartH} and quantize the pump field by replacing
	\begin{align}
	 |\beta_\text{cl}|e^{-i \omega_L t } \rightarrow \hat b,
	\end{align}
	 where $\hat b$ is the annihilation operator of the pump quanta in the Schr\"odinger picture.
	 The semiclassical time-dependence is thus transformed into the free evolution of the operator $\hat b$.
	 In practice this can be realized via QED with a trapped ion (for theory and corresponding experiments, see~\cite{Fid02a,Fid02b} 
	 and~\cite{Blatt02,Walther04}, respectively).
	 
	 The total Hamiltonian $ \hat{\mathcal{H}}$, in the Schr\"odinger picture, including the quantized pump field, reads as
	 \begin{align}
	 \hat{\mathcal{H}}=& \underbrace{\left(  \hbar \nu \hat a^\dag \hat a  +\hbar \omega_L \hat b^\dag \hat b+ \hbar \omega_{21} \hat A_{22}\right)}_{=\hat{\mathcal{H}}_0}   \label{Eq:H0}    \\
	 &+ \underbrace{\left( \hbar  \kappa \hat A_{21} \hat b \hat g( \eta(\hat a + \hat a^\dag)) + \text{H.c.} \right)}_{=\hat{\mathcal{H}}_\text{int}},  \label{Eq:Hinteraction1}
	 \end{align}
	 which is now time independent. As in the semiclassical case [see Eq.~\eqref{Eq:StartH}], this Hamiltonian is again based on the optical rotating-wave approximation.
	 The modified free Hamiltonian, $\hat{\mathcal{H}}_0$, now includes the free evolution of the quantized pump field with the frequency $\omega_L$.
	 Here, we again assume that we operate in the resolved sideband regime and quasiresonantly drive the $|1,n \rangle \leftrightarrow |2,n-k\rangle$ transitions in the vibronic rotating-wave approximation.
	 As before, only those terms of $\hat g$ are relevant which belong to this transition [see Eq.~\eqref{Eq:Expansiong}]:
	\begin{align}
	 \hat g( \eta(\hat a + \hat a^\dag))\rightarrow \hat f_k(\hat a^\dag \hat a;\eta) \hat a^k,
	\end{align}
	with $\hat f_k$ being defined in Eq.~\eqref{Eq:def-f}.
	Thus, we arrive at 
	 \begin{align}
	 \label{Eq:Hinteraction2}
	  \hat{\mathcal H}_\text{int}  = \hbar  \kappa \hat A_{21} \hat b \hat f_k(\hat a^\dag \hat a;\eta) \hat a^k + \text{H.c.}
	 \end{align}
	The interpretation of the Hamiltonian~\eqref{Eq:Hinteraction2} is the following:
	A pump photon is absorbed ($\hat b$) and the ion is excited ($\hat A_{21}$). 
	The vibrational transitions ($\hat f_k(\hat a^\dag \hat a;\eta) \hat a^k $) occur according to our chosen quasiresonance condition, $|n\rangle \rightarrow |n-k\rangle$. 
	The $\text{H.c.}$ term in addition describes the emission of a pump photon ($\hat b^\dag$), accompanied by the electronic transition $|2\rangle \rightarrow |1\rangle$ and the vibrational transition $|n-k\rangle \rightarrow |n\rangle$.
	As before, the pump field is not exactly on resonance, as $\Delta \omega \neq  0$. 
	In the case of interest, $\Delta \omega \ll \nu$, only the wanted transitions significantly contribute to the dynamics.
	Since the resulting Hamiltonian is not explicitly time dependent anymore, we obtain the time-evolution operator in the form
	\begin{align}
	 \label{Eq:timevolutionoperator}
	\hat{\mathcal{U}}(t) = \exp \left\{ - \frac{i (t-t_0)}{\hbar} \left[ \hat{\mathcal{H}}_0+ \hat{\mathcal H}_\text{int} \right]\right\},
	\end{align}
	with the definitions of the Hamiltonian according to Eqs.~\eqref{Eq:H0} and~\eqref{Eq:Hinteraction2}.
	
        \subsection{Solution for the quantized pump field}
	
	In this section we solve the time-evolution problem based on the operator~\eqref{Eq:timevolutionoperator}, i.e., we derive an analytical expression for $\hat{\mathcal{U}}(t)$.
	Our full Hamiltonian [see Eqs.~\eqref{Eq:H0} and~\eqref{Eq:Hinteraction2}] reads as
	\begin{align}
	\hat{\mathcal{H}}=& \left(  \hbar \nu \hat a^\dag \hat a  +\hbar \omega_L \hat b^\dag \hat b+ \hbar \omega_{21} \hat A_{22}\right) \nonumber \\
	 &+ \left(\hbar \kappa \hat A_{21} \hat b \hat f_k(\hat a^\dag \hat a;\eta) \hat a^k  + \text{H.c.} \right).
	\end{align}
	The eigenstates of this Hamiltonian are 
	\begin{align}
	|\psi^\pm_{mn} \rangle=c^\pm_{mn}( |2,m,n\rangle + \alpha^\pm_{mn} |1,m+1,n+k \rangle ),
	\end{align}
	where $|i,m,n\rangle$ denotes the electronic, pump-photon ($m=0,1,2,\dots$), and motional excitations. 
	The normalization yields $c^\pm_{mn}=\frac{1}{\sqrt{1+|\alpha^\pm_{mn} |^2}}$.
	The general procedure resembles that in Sec.~\ref{SubSec:TimeOrdering}. However, now we have an additional mode and we will solve the problem in the Schr\"odinger picture.
	
	The parameters are found to be
	\begin{align}
	 & \alpha_{mn}^\pm =\frac{\Delta \omega \pm \sqrt{\Delta \omega^2 + |\Omega_{mn}|^2 }}{ \Omega_{mn} },   \nonumber \\
	 & \omega_{mn}^\pm=  \frac{1}{2} \{ \Delta \omega ( 2m+1) + \nu ( 2n-2km) + \omega_{21} (2m+2)  \nonumber \\
	   & \pm \sqrt{\Delta \omega^2 + |\Omega_{mn} |^2} \},
	\end{align}
	where we used Eq.~\eqref{Eq:mismatch-cond} and defined
	\begin{align}
	\label{Eq:Rabi}
	  \Omega_{mn} = 2 \kappa \sqrt{m+1} f_k(n;\eta)\sqrt{\frac{(n+k)!}{n!}}.
	\end{align}
	Here, $ \omega_{mn}^\pm$ are the eigenvalues of $\hat{\mathcal H}$, associated with the eigenstates $|\psi_{mn}^\pm \rangle$, and $\Omega_{mn}$ is the nonlinear $k$-quantum Rabi frequency, which was already discussed in Ref.~\cite{V95}. 
	For the present problem there  occurs in Eq.~\eqref{Eq:Rabi} the additional factor $\sqrt{m+1}$, which is caused by the quantum treatment of the pump field.

	The completeness relation reads
	\begin{align}
	 &\hat 1 = \sum_{\sigma=\pm} \sum_{m,n=0}^\infty |\psi_{mn}^\sigma\rangle \langle \psi_{mn}^\sigma | + \sum_{n=0}^\infty |1,0,n\rangle \langle 1,0,n | \nonumber \\
	 & + \sum_{m=0}^\infty \sum_{q=0}^{k-1} |1,m+1,q \rangle \langle 1,m+1,q|.
	\end{align}
	Using the latter, we can rewrite the full time-evolution operator, Eq.~\eqref{Eq:timevolutionoperator}, in the form
	\begin{align}
	& \hat{\mathcal{U}}(t) = \sum_{\sigma=\pm} \sum_{m,n=0}^\infty e^{-i \omega_{mn}^\sigma \Delta t }|\psi_{mn}^\sigma \rangle \langle \psi_{mn}^\sigma | \nonumber \\
	 &+\sum_{n=0}^\infty e^{-i\nu n \Delta t} |1,0,n\rangle \langle 1,0,n| \nonumber \\
	 & + \sum_{m=0}^\infty \sum_{q=0}^{k-1} e^{-i [\nu q + \omega_L(m+1) ] \Delta t} |1,m+1,q\rangle \langle1,m+1,q|,
	\end{align}
	with $\Delta t=t-t_0$.
	For convenience we use the scaled (dimensionless) parameters:
	\begin{align}
	 &\Delta\tilde t := |\kappa| \Delta t ,  \nonumber \\
	 & \tilde \nu := \nu/|\kappa|  , \nonumber \\
	 & \tilde \omega_{21,L} := \omega_{21,L}/|\kappa| , \nonumber \\
	 &\tilde \Omega_{mn} := \Omega_{mn}/|\kappa|  , \nonumber \\
	 &\Delta \tilde \omega := \Delta \omega/ |\kappa| ,
	\end{align}
	with $\Delta\tilde t=\tilde t - \tilde t_0 $.
	In terms of these dimensionless quantities the unitary time-evolution operator reads as
	\begin{align}
	\label{Eq:FullSolution}
	& \hat{\mathcal{U}}(\tilde t) = \sum_{\sigma=\pm} \sum_{m,n=0}^\infty e^{-i \tilde \omega_{mn}^\sigma \Delta\tilde t}|\psi_{mn}^\sigma \rangle \langle \psi_{mn}^\sigma | \nonumber \\
	 &+\sum_{n=0}^\infty e^{-i \tilde \nu n \Delta\tilde t} |1,0,n\rangle \langle 1,0,n| \nonumber \\
	 & + \sum_{m=0}^\infty \sum_{q=0}^{k-1} e^{-i [\tilde \nu q + \tilde \omega_L(m+1) ]\Delta\tilde t } |1,m+1,q\rangle \langle1,m+1,q|,
	\end{align}
	where
	\begin{align}
	\label{Eq:rescaledQuant}
	 & \tilde \omega_{mn}^\pm =  \frac{1}{2} \{ \Delta \tilde \omega ( 2m+1) + \tilde \nu ( 2n-2km) + \tilde \omega_{21} (2m+2)  \nonumber \\
	 &  \pm \sqrt{[\Delta \tilde \omega]^2 + | \tilde \Omega_{mn} |^2} \}.
	\end{align}
	
	\subsection{Semiclassical versus quantized pump}
	\label{Subsec:Validity}
	Let us now compare the analytical results obtained from the solution of the quantized-pump dynamics with the numerical solutions (see Sec.~\ref{SubSec:TimeOrdering}) for a classical pump field, when the Hamiltonian is explicitly time dependent [see Eq.~\eqref{Eq:Hinteraction2-intpic}].
	As an example, we calculate the occupation probability of the excited electronic state:
	\begin{align}
	 \sigma_{22}(\tilde t) = \sum_{m,n} \langle 2,n,m|  \hat \rho(\tilde t) |2,n,m \rangle
	\end{align}
	where $\hat \rho(\tilde t) = \hat{\mathcal{U}}(\tilde t) \hat \rho(0) \hat{\mathcal{U}}(\tilde t)^\dag$ is the full density matrix of the state.
	\begin{figure}[h]
	\centering
	\includegraphics*[width=8.0cm]{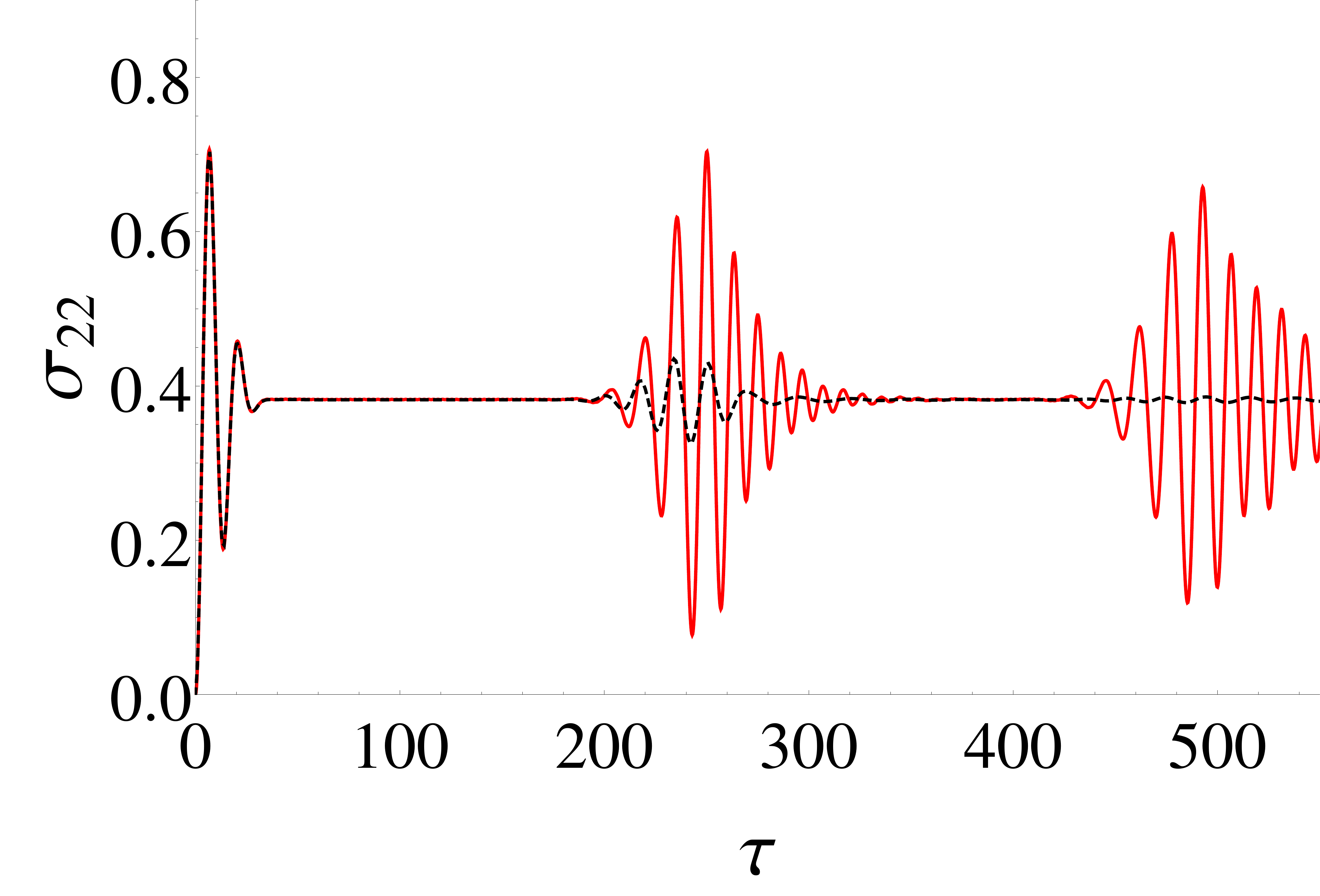}
	\caption{Comparison of the numerical solution for the classical pump, using the Hamiltonian~\eqref{Eq:Hinteraction2-intpic} in Eq.~\eqref{Eq:U1} (solid, red line) for $r=0.2$ 
	[see Eq.~\eqref{Eq:dimquant1}], with the analytical solution for the quantized pump field for $\beta_0=20$ (dashed, black line). 
	To obtain an equal time scaling of the semiclassical and quantum results, the latter solution was adjusted according to $\Delta \tilde \omega = |\beta_0| r = 4 $ [see Eq.~\eqref{Eq:dimquant1}] and $\tilde t = \tau/|\beta_0|$ [see Eq.~\eqref{Eq:dimquant2}].
	The initial state of the quantized motion was assumed to be a coherent state $|\alpha_0 \rangle$.
	Parameters: $\alpha_0=\sqrt{12}$, $k=2$, $\Delta \phi=0$, and $\eta=0.2$. 
	}\label{Fig:CQpump1}
	\end{figure}
	Using the input state $\hat \rho(0)=|1,\beta_0,\alpha_0\rangle \langle 1,\beta_0,\alpha_0 |$ at $\tilde t_0=0$, the analytical treatment yields
	\begin{align}
	\label{Eq:sigma22-quantized}
	& \sigma_{22}(\tilde t)=\sum_{m,n=0}^\infty \sum_{\sigma,\sigma'=\pm} e^{i[\tilde \omega_{mn}^{\sigma'}-\tilde \omega_{mn}^{\sigma}] \tilde t}  |c_{mn}^\sigma c_{mn}^{\sigma'}|^2  \nonumber \\
	 & \times ( \alpha_{mn}^{\sigma} )^\ast \alpha_{mn}^{\sigma'} 
	 \frac{|\beta_0|^{(2m+2)} |\alpha_0|^{(2n+2k)}}{(m+1)!(n+k)!} e^{-|\beta_0|^2-|\alpha_0|^2}.
	\end{align}
	Note that, due to the dependence on ${\tilde \omega_{mn}^{\sigma'} -\tilde \omega_{mn}^{\sigma}}$, there is no dependence of $ \sigma_{22}(\tilde t)$ on $\tilde \nu$ and $\tilde \omega_{21}$.
	\begin{figure}[b]
	\centering
	\includegraphics*[width=8.0cm]{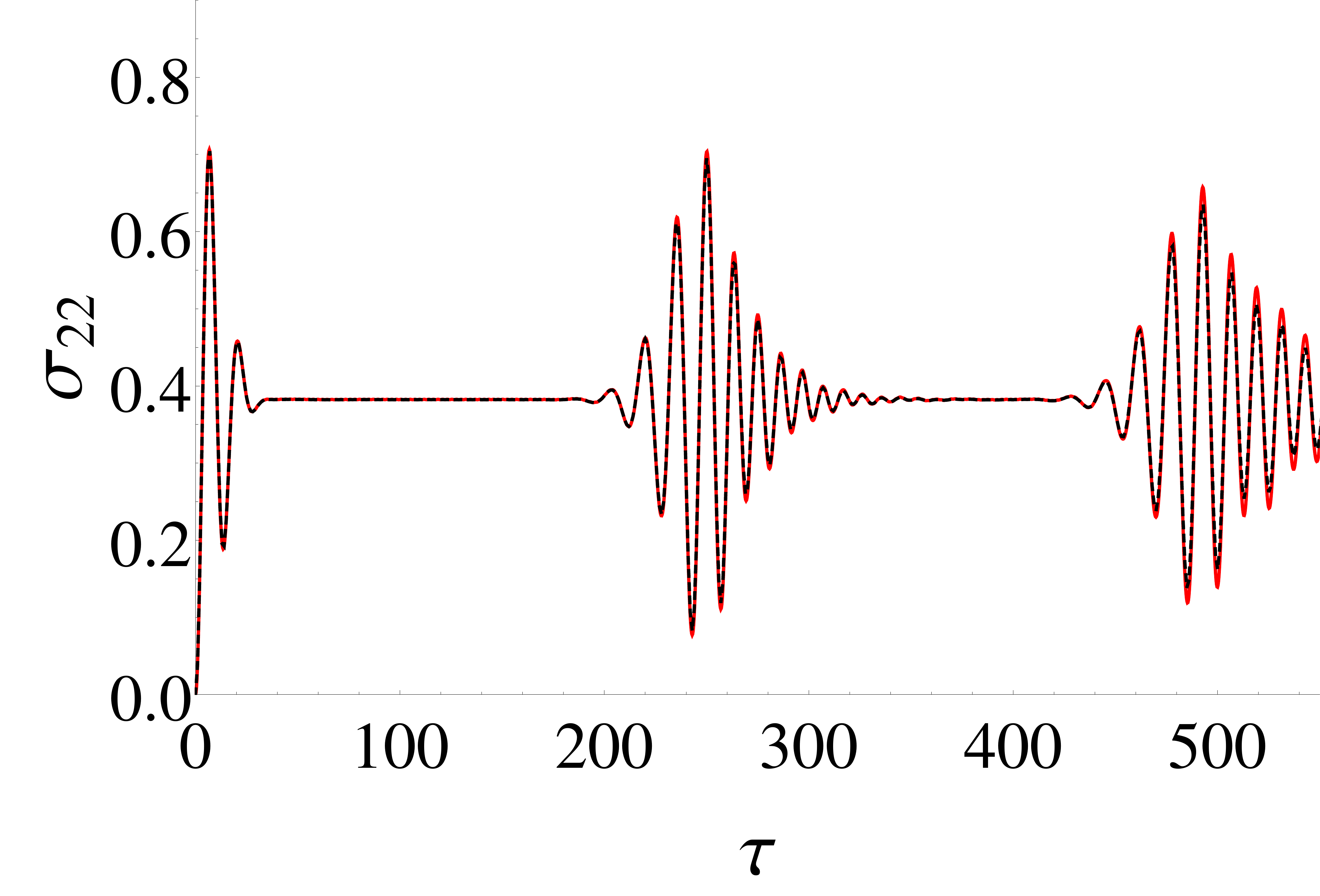}
	\caption{The same as in Fig.~\ref{Fig:CQpump1}, but for a strong quantized pump with $\beta_0 =100$ and $\Delta \tilde \omega = |\beta_0| r = 20 $.
	}\label{Fig:CQpump2}
	\end{figure}
	
	Let us consider the evolution for a relatively weak pump amplitude (see Fig.~\ref{Fig:CQpump1}).
	We see that, excluding the short-time dynamics, the solutions with classical and quantized pump differ significantly from each other. Hence, the used pump amplitude is by far not sufficiently large to be referred to as ``quasiclassical.''
	
	In the case of a stronger pump field (see Fig.~\ref{Fig:CQpump2}), we indeed obtain a dynamics which is almost identical to the numerical solution for a classical pump field.
	This not only enables us to conclude which pump amplitudes are needed such that the pump can be treated as a classical one on the corresponding time scale
	but also reveals that the solution found via quantization of the pump yields a more general description of the quantum system under study, where the time ordering is contained via the extension of the Hilbert space.

\section{Evolution of nonclassicality}\label{Sec:Pfunc}
	
	In this section we use the solution Eq.~\eqref{Eq:FullSolution}, to discuss the nonclassical properties of the system.
	Let us first introduce the notion of nonclassicality to be used in the following.
	
	\subsection{The regularized $P$~function}
	
	Using the Glauber-Sudarshan $P$~function~\cite{S63,G63}, $P(\alpha;t)$, any quantum state, given by its density operator $\hat \rho(t)$ at time $t$, can be expressed as a mixture of coherent states $|\alpha\rangle$:
	\begin{align}
	\hat \rho (t) = \int d^2\alpha P(\alpha;t) |\alpha \rangle \langle \alpha |.
	\end{align}
	We call a state \textit{nonclassical} if it cannot be expressed as a classical mixture of coherent states.
	In such cases $P(\alpha;t)$ cannot be interpreted in terms of a classical probability density~\cite{TG65,M86}, i.e., it can attain negative values in the sense of distributions.
	However, for many states $P(\alpha;t)$ is highly singular and, hence, it is not accessible in experiments.
	To uncover the negativities of $P(\alpha;t)$ it is therefore necessary to use a regularization procedure which yields a regularized version of this function~\cite{Kiesel10}.
	This procedure was successfully applied to experimental data~\cite{Bellini11,Kiesel11,Agu15} and generalized to different scenarios~\cite{Agu13,Agu17,K17}.
	Here we will only recapitulate the basic idea.
	
	The $P$~function is defined by the Fourier transform of the characteristic function ${\Phi(\beta;t)=\langle \hat D(\beta;t) \rangle e^{|\beta|^2/2} }$ with
	$\hat D(\beta;t) =e^{\beta \hat a(t)^\dag - \beta^\ast \hat a(t) }$:
	\begin{align}
	\label{Eq:DefP}
	P(\alpha;t)= \pi^{-2} \int d^2\beta e^{\beta^\ast \alpha - \beta \alpha^\ast} \Phi(\beta;t).
	\end{align}
	The possibly occurring singular behavior of $P$ results from the fact that $\Phi(\beta;t)$ may be unbounded and, hence, not square-integrable.
	According to Eq.~\eqref{Eq:DefP}, the $P$~function can therefore be highly singular.
	To get experimental access to the latter, one may introduce a filter function $\Omega_w(\beta)$ with some filter width $w$, to define the regularized $P$~function~\cite{Kiesel10} as
	\begin{align}
	\label{Eq.DefPregul}
	P_\Omega(\alpha;t)=\pi^{-2} \int d^2\beta e^{\beta^\ast \alpha - \beta \alpha^\ast}  \Omega_w(\beta)  \Phi(\beta;t)  .
	\end{align}

	The resulting function $P_\Omega(\alpha;t)$ is a regular and smooth~\cite{Agu13} function as long as the following requirements to the filter function are fulfilled.
	\begin{enumerate}
		\item $\Omega_w(\beta) \Phi(\beta;t)$ can be Fourier transformed for all filter widths $w$, with $w < \infty$.
		\item The Fourier transform of ${\Omega_w}$ is a probability density, so that it is non-negative.
		\item For a filter which is infinitely broad, $w \rightarrow \infty$, we obtain the original $P$~function, $P_\Omega \rightarrow P$.
	\end{enumerate}
	For an overview and the discussion of different filter functions we refer to Ref.~\cite{BK14}.
	
	\subsection{Calculation of $P_\Omega$ in Fock basis}
	
	In practical calculations, to obtain the full information on the quantum state, the density matrix of the state is calculated.
	In the following, we implement a suitable procedure to calculate $P_\Omega(\alpha;t)$ directly out of $\hat \rho(t)$.
	Let us first rewrite the definition of $P_\Omega$ [see Eq.~\eqref{Eq.DefPregul}] in Fock state basis:
	\begin{align}
	\label{Eq:PregulFock1}
	&P_\Omega(\alpha;t) \nonumber \\
	&=\pi^{-2} \int d^2\beta e^{\beta^\ast \alpha - \beta \alpha^\ast}  \Omega_w(\beta)  e^{|\beta|^2/2} \text{Tr} \left\{ \hat \rho (t) \hat D(\beta;0) \right\}  \nonumber \\
	  &=\sum_{m,n=0}^\infty  \rho_{mn}(t) \underbrace{  \int \frac{d^2\beta}{\pi^2} e^{\beta^\ast \alpha - \beta \alpha^\ast +|\beta|^2/2}  \Omega_w(\beta)   D_{nm}(\beta;0)}_{:=P_{\Omega,nm}(\alpha)},
	\end{align}
	with $\rho_{mn}(t)=\langle m| \hat  \rho(t) | n \rangle$ and $D_{nm}(\beta)=\langle n| \hat D(\beta;0) | m \rangle$.
	The functions $P_{\Omega,nm}(\alpha)$ are the regularized elements of the $P$~function in the Fock basis.
	In Eq.~\eqref{Eq:PregulFock1}, the complete time evolution is contained in the density matrix elements $\rho_{mn}(t)$.
	The functions $P_{\Omega,nm}(\alpha)$ however, only depend on the fixed parameter $w$ and the phase-space coordinate $\alpha$.
	Hence, it is possible to calculate these elements only once and after that we apply them to $\hat \rho(t)$, for arbitrary $t$.
	Let us therefore find a suitable expression of $P_{\Omega,nm}(\alpha)$.
	
	We make use of~\cite{VW06}
	\begin{align}
	 &\langle n | \hat D(\beta;0) | m \rangle 
	 = e^{-|\beta|^2/2} e^{i \varphi_\beta (n-m)} \Lambda_{nm}(|\beta|),
	 \end{align}
	 with
	 \begin{align}
	 \Lambda_{nm}(|\beta|)=
	 \begin{cases}
	    (-|\beta|)^{m-n} \sqrt{\frac{n!}{m!}} L_n^{(m-n)}(|\beta|^2) &  m \geq n \\
	    |\beta|^{n-m} \sqrt{\frac{m!}{n!}} L_m^{(n-m)}(|\beta|^2) &  m < n ,
	 \end{cases}
	\end{align}
	where $ L_n^{(k)}(x)$ are the generalized Laguerre polynomials.
	Using a radial symmetric filter function~\cite{BK14}, $\Omega_w(\beta) \equiv \Omega_w(|\beta|)$ with $x=|x|e^{i \varphi_x}$ for $\alpha$ and $\beta$, then $P_{\Omega,nm}(\alpha)$ may be rewritten as 
	\begin{align}
	 \label{Eq:PregulFock2}
	& P_{\Omega,nm}(\alpha) = \pi^{-2}  \int_0^\infty d|\beta|  \Lambda_{nm}(|\beta|) \Omega_w(|\beta|) |\beta|  \nonumber \\
	& \times \int_0^{2 \pi} d \varphi_\beta \,  e^{2 i |\alpha \beta| \sin(\varphi_\alpha-\varphi_\beta)}  e^{i \varphi_\beta (n-m)} .
	\end{align}
	The phase integral can be evaluated via substitution of the limits of integration:
	\begin{align}
	  &\int_0^{2 \pi} d \varphi_\beta \,   e^{2 i |\alpha \beta| \sin(\varphi_\alpha-\varphi_\beta)}  e^{i \varphi_\beta (n-m)} \nonumber \\
	  &= (-1)^{n-m} \int_{-\pi}^\pi d\varphi e^{-2 i|\alpha \beta| \sin(\varphi_\alpha-\varphi)} e^{i \varphi (n-m)} \nonumber \\
	  & = 2 \pi  e^{i (n-m)\varphi_\alpha} J_{n-m}(2 |\alpha \beta|),
	\end{align}
	where $J_n(x)$ are the Bessel functions of the first kind.

	Finally, we arrive at the expression
	\begin{align}
	 \label{Eq:PregulFock3}
	& P_{\Omega,nm}(\alpha) = \frac{2}{\pi}  e^{i (n-m)\varphi_\alpha}  \int_0^\infty d|\beta|  \Lambda_{nm}(|\beta|) \Omega_w(|\beta|) |\beta|  \nonumber \\
	& \times  J_{n-m}(2 |\alpha \beta|).
	\end{align}
	This relation holds true for all radial symmetric filters $\Omega_w(|\beta|)$.
	We will use the filter~\cite{BK14}
	\begin{align}
	\label{Eq:BenjFilter}
	 \Omega_w(|\beta|)=\frac{2}{\pi} \left[ \arccos\left(\frac{|\beta|}{2w}\right) - \frac{|\beta|}{2w} \sqrt{1- \frac{|\beta|^2}{4w^2}}  \right] \text{rect}\left(\frac{|\beta|}{4w} \right),
	\end{align}
	with $\text{rect}(x)=1$ if $x \leq 1/2$ and $\text{rect}(x)=0$ elsewhere.
	Inserting Eq.~\eqref{Eq:BenjFilter} in Eq.~\eqref{Eq:PregulFock3} and using the substitution $z:= |\beta|/(2w)$ yields
	\begin{align}
	 \label{Eq:PregulFock4}
	&P_{\Omega,nm}(\alpha) = \frac{16}{\pi^2}w^2  e^{i (n-m)\varphi_\alpha}  \int_0^1 dz  \Lambda_{nm}(2wz)  z \nonumber \\
	&\times  J_{n-m}(4 w |\alpha|z) \left[ \arccos(z)-z\sqrt{1-z^2} \right]. 
	\end{align}
	The $z$ integral needs to be evaluated numerically in general.
	Note that here $P_{\Omega,nm}(\alpha)=P_{\Omega,mn}(\alpha)^\ast$ holds.
	We stress that this procedure applies to any time evolution.

	\subsection{Nonclassicality in the nonlinear Jaynes-Cummings model}
	
	We are interested in the nonclassical properties of the vibrational states.
	Hence we calculate the reduced density matrix:
	\begin{align}
	 \hat \rho_\text{vib}(\tilde t) = \sum_{i=1,2} \sum_{m=0}^\infty \langle i,m | \hat \rho(\tilde t) |i,m\rangle,
	\end{align}
	where the trace over the electronic states and the pump states is evaluated.
	Here we use $\hat \rho(\tilde t) = \hat{\mathcal{U}}(\tilde t) \hat \rho(0) \hat{\mathcal{U}}(\tilde t)^\dag$ with the time-evolution operator given in Eq.~\eqref{Eq:FullSolution}
	and $\hat \rho(0)=|2,\beta_0,\alpha_0\rangle \langle 2,\beta_0,\alpha_0 |$ at $\tilde t_0=0 $. 
	This yields
	\begin{align}
	& \hat \rho_\text{vib}(\tilde t) = \sum_{\sigma,\sigma'=\pm} \sum_{\substack{n,n'=0\\m=0}}^\infty e^{i [\tilde \omega_{mn'}^{\sigma'} - \tilde \omega_{mn}^\sigma] \tilde t} 
	 |c_{mn}^\sigma c_{mn'}^{\sigma'}|^2 \nonumber \\
	 &\times  \frac{|\beta_0|^{2m}e^{-|\beta_0|^2}}{m!} \frac{\alpha_0^n \alpha_0^{\ast n'} e^{-|\alpha_0|^2}}{\sqrt{n! n'!}} \nonumber \\
	 &\times \left\{ |n \rangle \langle n'| + \alpha_{mn}^\sigma ( \alpha_{mn'}^{\sigma'} )^\ast |n+k \rangle \langle n'+k|\right\},
	\end{align}
	which does not depend on $\tilde \omega_{21}$.
	
\begin{figure}[h]
	\centering
	\includegraphics*[width=8.0cm]{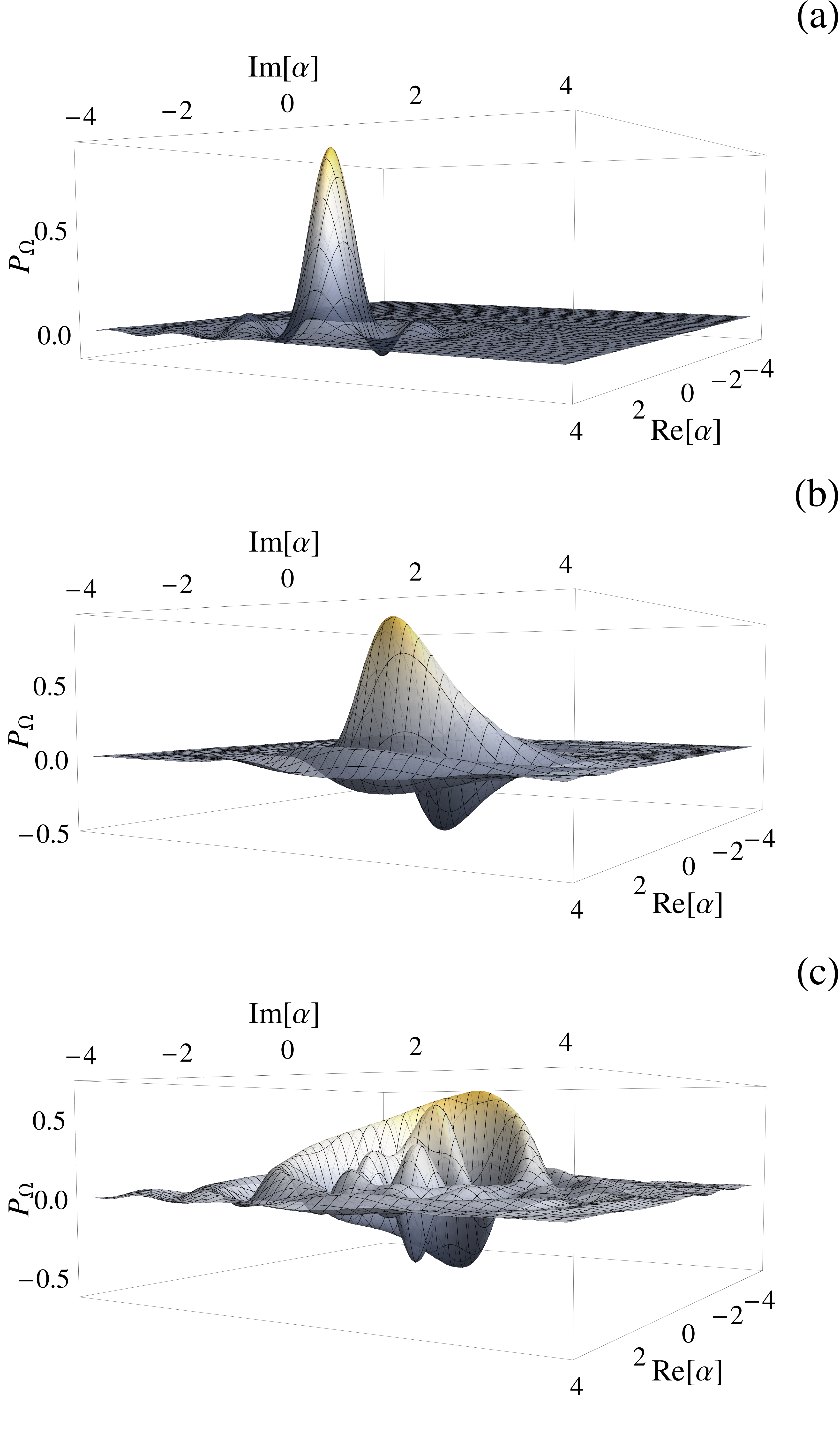}  
	\caption{The regularized Glauber-Sudarshan $P$~function, derived via Eq.~\eqref{Eq:PregulFock1} together with Eq.~\eqref{Eq:PregulFock4}, is shown for the initial coherent state $|\alpha_0\rangle$, at 
	$\tilde t=4\ (a),\ 13 \ (b),\ 50 \ (c)$. 
	Parameters: $\alpha_0=\sqrt{5}$, $k=3$, $\Delta \phi=\pi/2$, $\eta=0.2$, $\tilde \nu=5000$, $\beta_0=40$, $\Delta \tilde \omega=8$, and $w=1.7$.
	}\label{Fig:3dP}
\end{figure}
	The surface plot of the regularized $P$~function is given in Fig.~\ref{Fig:3dP}.
	Due to the vibronic coupling the initial motional coherent state at $\tilde t=0$ evolves into a nonclassical state.
	For a rather small time [see Fig.~\ref{Fig:3dP} (a)], the state is still close to a coherent one. For larger times, one obtains more distorted states [see Figs.~\ref{Fig:3dP} (b) and (c)]. 
	The quantum character is displayed by the clearly visible negativities of the regularized $P$~functions at the corresponding times.
	The choice of $\tilde \nu$ does not affect the nonclassical properties of the state but leads to a rotation in phase space. 
	We note that the nonclassical effects shown in Fig.~\ref{Fig:3dP} become smaller for increasing frequency mismatch $\Delta \tilde \omega$, as the nonlinear vibronic interaction becomes weaker in this case.
	
	Finally we would like to note that the nonclassicality quasiprobabilities shown in Fig.~\ref{Fig:3dP} can be  determined straightforwardly in experiments. 
	This can be done by the method introduced in~\cite{WaVo95} and realized in~\cite{Monroe05,Monroe05b}, which allows the direct measurement of the characteristic function $\Phi(\beta, t)$ of the $P$~function [see Eq.~\eqref{Eq:DefP}]. 
	This technique can be readily combined with the direct sampling approaches as developed for the nonclassicality quasiprobabilities of radiation fields~\cite{Kiesel11,Agu15}.
	
\section{Summary and Conclusions}\label{Sec:Conclusions}

	In this paper we considered the time-dependent Hamiltonian of a nonlinear Jaynes-Cummings system that is driven in quasiresonance.
	We showed, that time-ordering effects have a crucial impact on the system and can therefore not be omitted.
	As the general solution of a time-dependent Hamiltonian can become a cumbersome task, we introduce a method to circumvent this issue via quantizing the pump field.
	By extending the Hilbert space of the system, the dynamics becomes exactly solvable.
	Using the resulting time-independent Hamiltonian we derived an analytical expression of the time-evolution operator.
	
	For a pump field prepared in a coherent state the solutions were shown to converge to the classical-pump scenario where the discrepancies shrink with increasing coherent amplitude.
	Furthermore, we visualized the temporal evolution of the nonclassicality quasiprobability of the motional states of the ion.
	This regularized version of the often strongly singular Glauber-Sudarshan $P$~function has the advantage that it can be determined in experiments. 
	Their negativities certify the quantum nature of the system under study.
	The introduced method to calculate this quasiprobability out of the density matrix applies to any time evolution.
	
	In general, the derived algebra of the quasiresonantly driven trapped ion renders it possible to investigate complex scenarios where the interaction of the vibrational and the atomic (source) degrees of freedom is of interest.
	This may include the study of time-dependent motional quantum correlation effects. 
	Furthermore, our analytical approach may yield a deeper insight into the properties of non-equal-time commutation rules, in cases with explicitly time-dependent interactions.

\begin{acknowledgments}
	We thank Ruynet Lima de Matos Filho for helpful comments.
	W. V. acknowledges funding from the European Union's Horizon 2020 research and innovation programme under Grant No. 665148.
\end{acknowledgments}

\end{document}